# PROMOTING AI COMPETENCIES FOR MEDICAL STUDENTS: A SCOPING REVIEW ON FRAMEWORKS, PROGRAMS, AND TOOLS

YINGBO MA[1], YUKYEONG SONG[2], JEREMY A. BALCH[3], YUANFANG REN[1], DIVYA VELLANKI[1], ZHENHONG HU[1], MEGHAN BRENNAN[4], SURAJ KOLLA[1], ZIYUAN GUAN[1], BROOKE ARMFIELD[1], TEZCAN OZRAZGAT-BASLANTI[1], PARISA RASHIDI[5], TYLER J. LOFTUS[3], AZRA BIHORAC[†1], AND BENJAMIN SHICKEL[†1]

Abstract. As more clinical workflows continue to be augmented by artificial intelligence (AI), AI literacy among physicians will become a critical requirement for ensuring safe and ethical AI-enabled patient care. Despite the evolving importance of AI in healthcare, the extent to which it has been adopted into traditional and often-overloaded medical curricula is currently unknown. In a scoping review of 1,699 articles published between January 2016 and June 2024, we identified 18 studies which propose guiding frameworks, and 11 studies documenting real-world instruction, centered around the integration of AI into medical education. We found that comprehensive guidelines will require greater clinical relevance and personalization to suit medical student interests and career trajectories. Current efforts highlight discrepancies in the teaching guidelines, emphasizing AI evaluation and ethics over technical topics such as data science and coding. Additionally, we identified several challenges associated with integrating AI training into the medical education program, including a lack of guidelines to define medical students' AI literacy, a perceived lack of proven clinical value, and a scarcity of qualified instructors. With this knowledge, we propose an AI literacy framework to define competencies for medical students. To prioritize relevant and personalized AI education, we categorize literacy into four dimensions: *Foundational*, *Practical*, *Experimental*, and *Ethical*, with tailored learning objectives to the pre-clinical, clinical, and clinical research stages of medical education. This review provides a road map for developing practical and relevant education strategies for building an AI-competent healthcare workforce.

## Introduction

Emerging innovations in artificial intelligence (AI) have the potential to transform medical practice and healthcare delivery [34]. As of May 2024, the U.S. Food and Drug Administration (FDA) has authorized 882 AI-enabled medical devices [12]; in 2023, the European Medicines Agency (EMA) has also released a preliminary document outlining its view on the application of AI and machine learning (ML) throughout different stages of a medicine's life cycle [2]. It is anticipated that AI-powered medical devices will increasingly become integrated into

[1] Department of Medicine, University of Florida; [2] School of Teaching and Learning, University of Florida; [3] Department of Surgery, University of Florida; [4] Department of Anesthesiology, University of Florida; [5] Department of Biomedical Engineering, University of Florida;

yingbo.ma@ufl.edu, y.song1@ufl.edu, balch.jeremy@ufl.edu, renyuanfang@ufl.edu, dvellanki@ufl.edu, hzhuf@ufl.edu, megbren@ufl.edu, n.kolla@ufl.edu, ziyuan.guan@ufl.edu, barmfield@ufl.edu, tezcan@ufl.edu, parisa.rashidi@ufl.edu, tloftus@ufl.edu, abihorac@ufl.edu, shickelb@ufl.edu

† Authors contributed equally

Corresponding author: abihorac@ufl.edu

various areas of healthcare and clinical diagnostics [9, 4]. With the notable trend of increasing integration of AI devices into clinical workflows, medical students (i.e., future physicians) will play a crucial role in the utilization and management of medical AI systems [18]. Therefore, it is essential for medical educators and institutions to provide AI education and training to prepare medical students to adapt to the evolving healthcare environment.

Despite this prevailing trend, medical students are not adequately equipped with AI competencies [48, 13]. National surveys conducted in Canada in 2019 and in the U.S. in 2022 both revealed that fear towards AI is pervasive amongst medical students due to a lack of understanding, uncertainty about ethical and legal considerations surrounding the use of AI, and job security [43]. Establishing a clear framework of AI literacy for medical students remains an important goal for researchers and education practitioners [22, 28]. However, integrating AI-related courses into the loaded medical training curriculum is difficult and faces resistance largely due to the lack of consensus on how to incorporate change and an already busy curriculum [15, 42]. Most existing AI literacy frameworks are defined for K-12 students or the general public [31] and are not suitable for addressing the specific needs and requirements of medical students. While some have proposed sets of AI topics most relevant for medical students [27, 32, 37], there is little consensus and best practices for AI education in medical education remains unclear.

Most existing literature in this field addresses general perspectives on the use of artificial intelligence in medical education, with less focus on specific teaching frameworks and educational programs. In this comprehensive scoping review, we determined the current state of knowledge, identified gaps in AI education for medical students, and synthesized several guidelines and recommendations for developing AI-integrated medical curricula. We examined literature published from 2016-2024 to investigate the following two questions:

(1) What are the existing theoretical proposals and guidelines for teaching AI to medical students?
(2) What are the existing practical teachings to medical students?

In Question 1, we reviewed existing work that proposed guidelines for educating medical students about AI, with the goal of proposing a distilled comprehensive AI literacy framework for medical students, defining the necessary competencies and practical skills of safely and effectively using AI in various clinical settings. In Question 2, we reviewed current educational practices for providing AI training to medical students, including formal curricular implementations, informal extracurricular programs, and educational tools. Taken together, we sought to identify (1) discrepancies between existing theoretical AI education guidelines and existing education practice for medical students, (2) challenges of integrating AI education into medical training, (3) a potential lack of tailored educational tools, and (4) opportunities for researchers and educational practitioners to provide more flexible, engaging, and personalized AI medical learning experience for medical students in the future. To our knowledge, this is the first review to comprehensively assess both theoretical guidelines and practical teaching efforts regarding AI education for medical students.

## Methods

This scoping review searched and selected eligible studies using the criteria, guided by the PRISMA [25] framework to ensure a comprehensive and transparent approach to the scoping review process, presented in Figure 1.

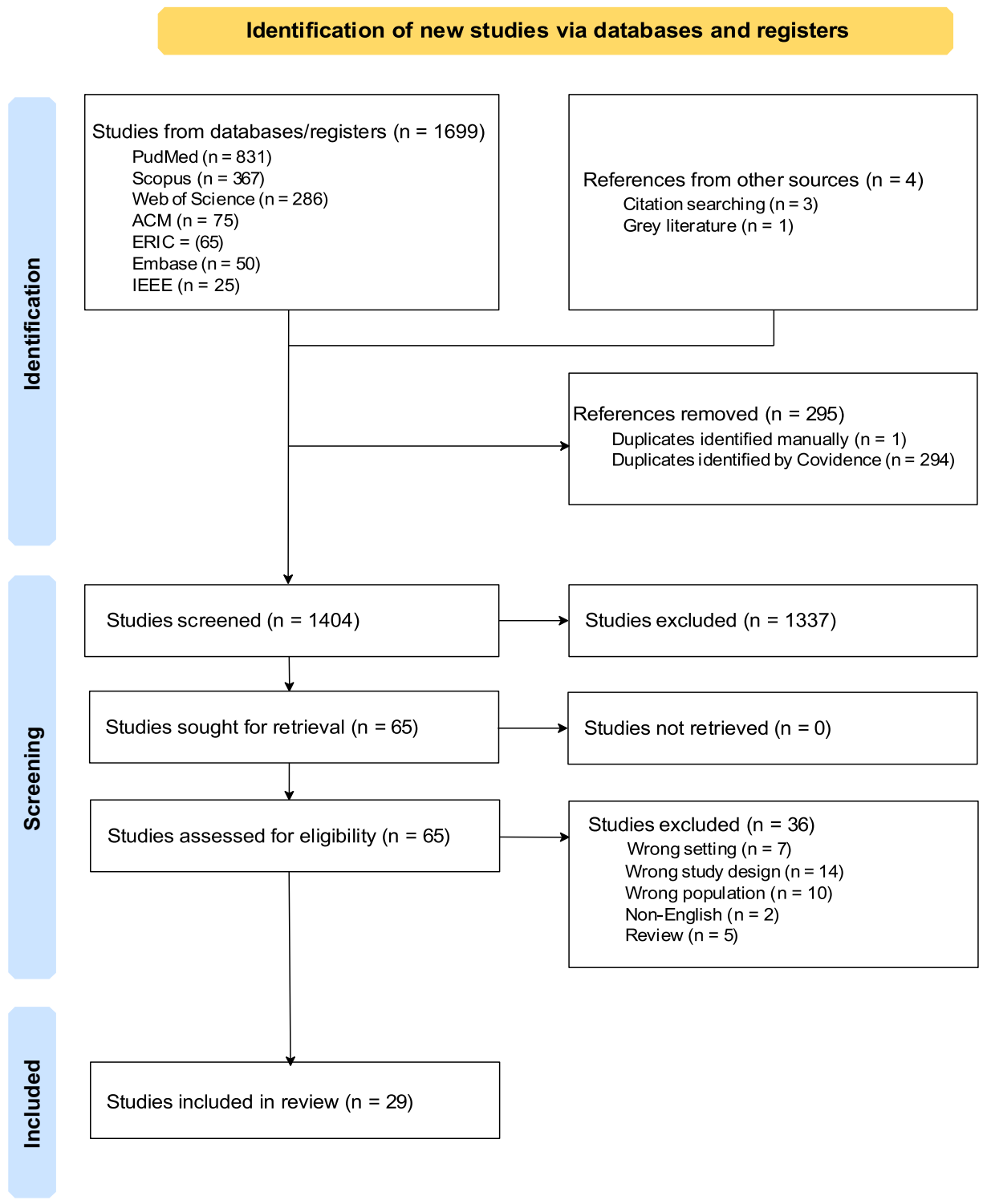

Figure 1. Search strategy and publication screening process: PRISMA flowchart.

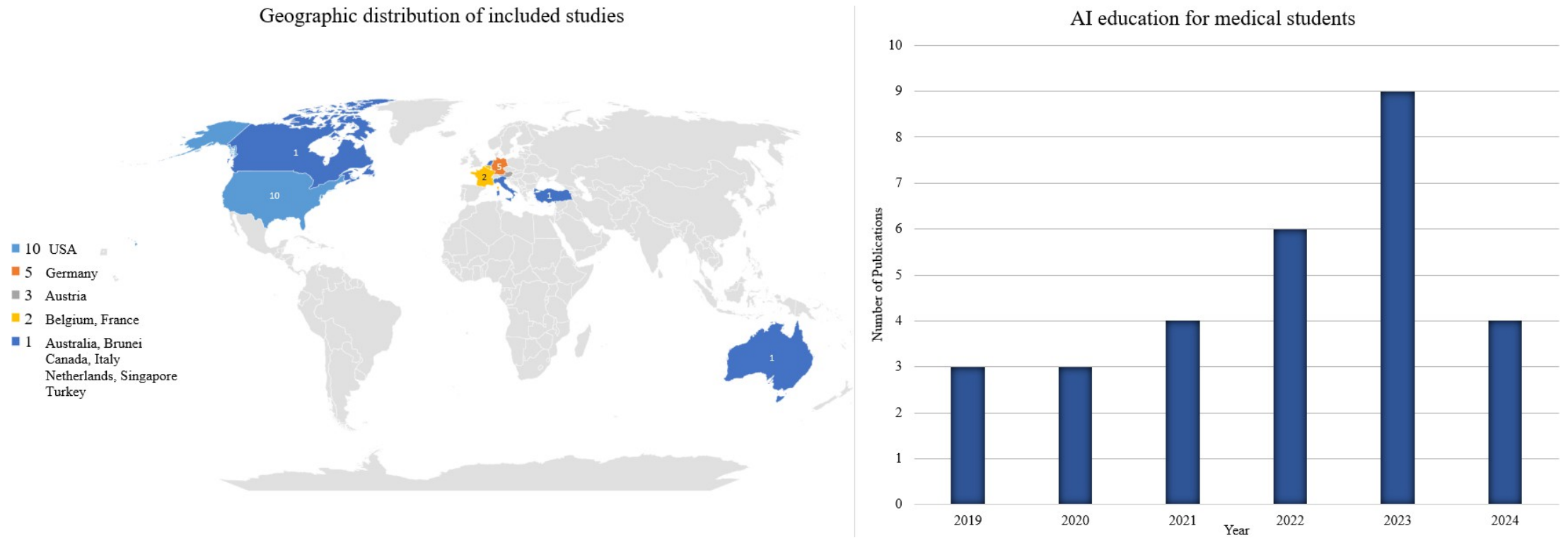

Figure 2. Distribution of included 29 studies

**Search strategy** Records were identified through database screening using search terms. We selected seven common medical and educational databases, including PubMed, Scopus, Web of Science, ACM digital library, Education Resources Information Center (ERIC) online library, Embase, and IEEE Xplore digital library. The search string was iteratively modified to ensure that the results specifically capture publications related to this scoping review. Following this iterative revision, our search string was:

> ("AI" OR "artificial intelligence" OR "machine learning" OR "deep learning" OR "natural language processing") AND ("teaching" OR “training” OR "curricul*" OR "program" OR "course") AND ("medical Student" OR "medical education" OR "medical training")

The asterisk (*) indicates a wildcard character that includes words of the same stem (i.e., both curriculum and curricula would be included in our search). We limited the period of publications starting from 2016, the year which began gaining general research interest in the field of AI education [46]. This search process resulted in 1,699 publications for screening.

**Screening** We imported the searched articles into the Covidence online platform for screening. The screening was performed in duplicate by two authors (Y.M. and S.K.) using a two-stage screening approach: title & abstract screening, followed by a full-text review. The study’s inclusion criteria focused on medical students, incorporating curricula that teach AI during formal and informal education, extracurricular programs, workshops, and online platforms like MOOCs. Excluded groups included post-graduate trainees, medical professionals, non-medical undergraduates, and K-12 students. Also excluded were papers using AI as learning tools or for evaluating performance. In the process, viewpoints, perspectives, or commentaries that provided detailed frameworks on AI education or medical students were included; articles that only emphasized the importance of AI education for medical students without proposing actionable guidelines were excluded; all review articles were also excluded but used as second-source. Finally, a total of 29 studies were included in this scoping review.

**Data extraction** For the 29 studies that fulfilled the inclusion criteria, four researchers extracted data in a standardized and predefined form independently and then combined the extracted information together to include comprehensive information and resolve any conflicts. For Question 1: What are the existing theoretical proposals and guidelines teaching AI to medical students, two authors (Y.M. and Y.S.) extracted and recorded title, year of publication, author, article type, and proposed teaching topics under the four key dimensions of the AI Literacy for the Public [31], including know and understand AI, use and apply AI, evaluate and create AI, and AI ethics. For Question 2: What are the existing practical teachings to medical students, another two authors (S.K., and D.V.) extracted and recorded title, year of publication, author, article type, program duration, program location, target audience, number of participants, program structure, AI topics covered, instructor expertise, education materials, education tools, assessments, key findings/outcomes, and program limitations (if reported).

## Results

A total of 29 studies fulfilled inclusion criteria. Figure 2 shows the number of publications per year, as well as the geographic distribution of the studies included. Among these 29 included studies, 18 studies suggested theoretical proposals, guidelines, or specific curriculum defining AI literacy for medical students (Question 1); 11 studies described formal coursework, programs, or workshops to provide AI training to medical students and assessed their learning (Question 2). Below, we summarize the results for each of our two questions.

---

https://www.covidence.org/

*Question 1: What should medical students know about AI: proposals and guidelines* In Question 1, we aimed to collect and synthesize the existing theoretical proposals and guidelines that define medical students' AI literacy. Medical students are both future "AI users" and "AI developers" [6]; thus, specifically tailored AI literacy frameworks will guide researchers and medical schools to provide medical students with relevant knowledge and skills to apply AI in their future careers. The data in 18 included studies that proposed theoretical guidelines were extracted and synthesized under the four key dimensions of the AI Literacy for the Public [31], a framework that defines the AI literacy for general public as four dimensions: know and understand AI, use and apply AI, evaluate and create AI, and AI ethics. The general characteristics of these 18 studies are shown in Table 1.

**Know and Understand AI**: Existing guidelines emphasize medical students' essential knowledge of AI without focus on AI related sub-fields such as probabilities, computer vision, and natural language processing. 17 out of 18 studies recommend medical students should a brief history of AI and ML [8], common AI terminologies and basic understanding of traditional ML algorithms (e.g., decision tree) [3], and what AI can do and cannot do for physicians [14, 16]. Eight out of 18 studies recommended more specific AI-related sub-fields as requirements, including basic concepts of data science [36, 27] and basic concepts of statistics [17, 8, 53]. Only two out of 18 studies recommended basic understanding of deep learning, neural networks, computer vision, and natural language process as requirements for medical students [47, 32].

**Use and Apply AI**: Existing guidelines emphasize practical use of AI in health data engineering nine out of 18 studies required medical students understand the important role of clinical data in AI innovation development [15], health data collection, curation, management, and protection [16, 32], and the whole health data processing pipeline for training ML models from data cleaning to feature engineering [27, 8]. 16 out of 18 studies specifically emphasized the importance of teaching medical students how to safely and effectively use clinical AI tools, including AI applications in general practices [36], different application scenarios of AI in medicine and the basic flow of how they work [15, 35, 16, 27], fundamental knowledge of translating AI concepts into clinical practice [47], applying predictive AI techniques for diagnosis and treatment [14, 52], and critical appraisal of clinical AI tools [24, 35].

**Evaluate and Create AI**: Existing guidelines place less emphasis on medical students' knowledge and skills of evaluating and creating AI applications. Seven out of 18 studies suggested that medical students should have basic skills to evaluate and interpret the performance of clinical AI models, including understanding of common evaluation metrics [32], understanding the results of AI models through explainability analysis (e.g., feature importance) [8], and performing statistical testing to compare across multiple AI models [44, 35, 50, 32, 52]. In addition, only one studies suggested that designing a ML algorithm as a requirement during AI training for medical students [50].

**AI Ethics**: Existing guidelines focus on the ethical issues of AI in clinical practice. All 18 studies emphasized on equipping medical students with understanding of the black-box nature of ML models, as well as sources of data and model bias, leading to the explainability and fairness issue of AI models. As future "users" of clinical AI applications, medical students need to know how these issues directly relate to core clinical values [27], and be fully transparent on informed consent when communicating with patients [33, 53], as well as legal compliance [36, 47, 16], social impacts [16], and accountability [8, 3, 32, 53]. Three studies also include that medical students should be aware of health data protection and privacy policies, and known and potential unknown risks of AI models [47, 52, 53].

*Question 2: Teaching AI to medical students: courses and programs* In Question 2, we aimed to summarize current AI education in medical school, either courses integrated into medical curriculum or informal training programs such as extra-curricular programs and workshops. The data in 11 included studies were extracted and synthesized into main categories including type (either mandatory course, elective course, or extra-curriculum program), number of participants, structure and teaching materials, AI topics covered, instructor expertise background, education tools used during the course or program, assessment, key findings and outcomes, and limitations. The general characteristics of these 11 studies are shown in Table 2.

**Type**: Current AI programs for medical students provide them maximum flexibility to tailor their education according to their interests and schedules. The majority (eight out of 11) of AI training programs for medical students are not mandatory and offered as elective courswork. Another two studies provided AI training to medical students through extra-curricula programs and workshops [45, 23]. Only 1 study reported the results from a new mandatory AI program implemented for the bachelor's degree in medicine.

**Structure and teaching materials**: The program structures vary based on the different durations of the program (from 16 hours to one year); most AI training programs started with learning sessions (this can be in-person lecturing [21, 54, 23, 49, 20], online live lecturing [45, 19, 38], and online self-study of pre-recorded videos and reading materials [22, 40, 1]). 5 out of 11 studies supplemented students' self-learning with guided community discussion [21, 45, 22, 19, 38]. In addition, three out of 11 studies required medical students to work on collaborative projects on building AI tools [23, 49, 38], engaging students in hands-on activities to reinforce their uptake and understanding of AI concepts.

**Instructor Expertise**: Most studies recruited instructors from university faculties and AI researchers. Only one study recruited a medical informatics teacher with several clinicians, ensuring that students received insights from both technical and clinical perspectives.

**AI Topics Covered**: The common AI topics that were covered in these AI training programs provided a comprehensive understanding of AI and the necessary skills to create and evaluate AI models yet placed less emphasis on critical skills to leverage AI technologies effectively and critically in medical practice and AI ethics. In terms of **Know and Understand AI**, six studies out of 11 taught introduction to AI and ML [21, 23, 19, 40, 1, 38], including the basics of AI and ML, including various techniques such as supervised/unsupervised learning, classification, and regression. Only one study [21] explicitly covered the statistics and mathematical foundations of AI by including basic knowledge of statistics and mathematical description in AI procedures. Three studies covered the topic of neural networks and deep learning [21, 1, 38], teaching the principles of neural networks and deep learning essential for understanding complex AI models. In terms of **Use and Apply AI**, two programs covered the health data engineering topics [54, 40], introducing medical image data and electronic health records (EHRs), and their transformative role in healthcare delivery and research. Seven programs covered the topics of understanding and using AI applications in medicine [21, 54, 22, 19, 49, 40, 20], covering various applications of AI in medicine, clinical decision support systems, innovations driven by AI, value-based care, and precision medicine. In terms of **Evaluate and Create AI**, only one program explicitly covered the topic of critically evaluating AI models, while three programs taught students basic programming skills for practical implementation of customized AI models. In terms of **AI Ethics**, six out of 11 programs provided students with the training topics of AI safety and ethics [21, 45, 23, 19, 49, 1], including regulatory policy over medical AI applications and ethical issues that concern the use of ML in health care.

**Education Tools**: Only three out of 11 studies reported using online educational tools and platforms to manage and enhance learning [22, 19, 49]. In addition, these tools are popular learning platforms for AI education for the general public (e.g., MOOC ), and their learning schedules and content are not specifically designed to meet the rigorous demands and specialized needs of medical students aiming to apply AI in clinical practice.

**Assessment**: The study assessments focused on three themes: medical students knowledge uptake and understanding about AI, interest and confidence in AI, and overall satisfaction toward the course/program. Six out of 11 studies used pre- and post-study AI konwledge assessment. Among these, five [21, 45, 22, 1, 20] used subjective self-reported survey (e.g., Medical AI Readiness Scale (MAIRS-MS) [17]) and only one study [19] used objective AI knowledge test). Two studies surveyed students' self-reported interest and confidence in AI [21, 1]. Four studies [54, 23, 49, 40] used post-survey for students' satisfaction toward the lesson content, format, and global experience.

**Key Findings/Outcomes**: The key findings across the 11 studies reveal that integrating AI education into medical curricula significantly enhanced students' knowledge, confidence, and interest in AI. Six out of 11 studies reported substantial increases in AI knowledge, both self-assessed and measured objectively [21, 45, 22, 19, 1, 20]. For instance, quiz scores improved dramatically [1], and pre- and post-study test comparisons showed significant gains [19]. two studies [21, 1] reported that students' interest in AI topics consistently grew, and the programs elicited great enthusiasm and strong engagement. Three studies reported that medical students' overall satisfaction with the AI courses was high, and they perceived the lesson contents and formats generally attractive [54, 49, 38].

**Main Limitations**: The main limitations reported across the studies include small sample sizes, overreliance on flawed metrics such as self-reported confidence assessments, and self-selection bias in elective enrollment. Five studies [45, 49, 40, 1, 20, 38] emphasized small sample size of participants, limiting the generalizability of their findings to other institutions and broader populations. Three studies [19, 49, 38] reported that training courses or programs being electives naturally attracted students already interested in AI, which limited the generalizability of the results to all medical students. Two studies [22, 1] emphasized that using self-reported questionnaires, such as the MAIRS-MS [17], rather than objective performance tests, could bias the results. One study [49] mentioned that the program was fully in-person and inflexible, with difficulties in finding instructors with appropriate backgrounds who possess expertise in both clinical medicine and medical informatics.

https://www.mooc.org/

Table 1: What should medical students know about AI? Existing AI topics proposed to teach

| **Author, Year; Article Type** | **Four Aspects of AI Literacy (Ng et al., 2021 [31]): (1) *Know and Understand AI*, (2) *Use and Apply AI*, (3) *Evaluate and Create AI*, (4) *AI Ethics*** | | | | | | |
|---|---|---|---|---|---|---|---|
| | ***Know and Understand AI*** | | ***Use and Apply AI*** | | ***Evaluate and Create AI*** | | ***AI Ethics*** |
| | **AI Basics** | **AI Sub-fields** | **Health Data Engineering** | **Use of AI Tools** | **Evaluate AI Tools** | **Create AI Tools** | |
| (Paranjape et al., 2019 [36]); Viewpoint | Essential AI Concepts | Common data science terminologies | Health data curation, quality provenance, integration, and governance | AI application in general practices; Various AI applications in clinical scenarios | | | AI law; AI policies regarding medicine domain |
| (Wartman et al., 2019 [51]); Research | Basic understanding of AI | | | Collaborating with and management of AI applications; Using probability-based AI tools in clinical decision making | | | AI ethics |
| (Sapci et al., 2020 [44]); Review | Appropriate ML knowledge | | | Applying predictive AI techniques for purposes such as improve the efficiency of patient care | Compare and evaluate AI tools | | Ethical implications of using AI tools |
| (McCoy et al., 2020 [27]); Comment | Basic concepts about AI | Basic concepts of data science | Pipeline of health data acquisition and cleaning; Health data visualization | Using common AI applications in the clinical domain; The future potential of AI applications in the clinical domain; When and how to choose from different clinical AI applications | | | fairness, accountability, transparency, and how they directly relate to core clinical values |
| (Karaca et al., 2021 [17]); Research | Basic AI terminologies; How AI systems are trained; Organize basic workflow in accordance with the logic of AI | Basic concepts of data science; Basic concepts of statistics | | Using AI tools effectively in healthcare delivery; Explaining AI technologies to patients; Choosing the proper AI tools for the problems that arise in healthcare settings | | | Use health data legally in accordance with ethical norms; Use AI technologies in accordance with regulations |

Continued on next page

Table 1: What should medical students know about AI? Existing AI topics proposed to teach (Continued)

| | | | | | | | |
|---|---|---|---|---|---|---|---|
| (Ngo et al., 2021 [33]); Perspective | The types of tasks that are amenable to AI; AI's impact on the role of physicians | | General pipeline health data engineering workflows | Modern use cases of AI tools in medicine domain | | | Informed Consent; Algorithm bias and limitation |
| (Blanc et al., 2022 [8]); Viewpoint | A brief history of AI/ML methods; Basic AI algorithms (e.g., decision tree, support vector machine) | Basic concepts of mathematics for ML (e.g., probabilistics, statistics) | Big data in biomedical (infrastructure, types, integration); Biomedical data preprocessing, dimensional reduction, feature selection | | Explainability (e.g., SHAP, feature importance) | | Accountability; Transparency; Fairness/Algorithmic discrimination |
| (Gray et al., 2022 [14]); Research | Basic use of algorithms in AI | | | Using AI-based diagnosis and treatment applications, targeted and focused development of a group of individuals rather than the whole workforce" | | | Ethical implications of AI tools |
| (Grunhut et al., 2022 [15]); Viewpoint | What AI can do and cannot do | | The important role of data in clinical AI innovation development | The breadth of clinical AI tools and the basic flow of how they work | | | AI ethics |
| (Liaw et al., 2022 [24]); Research | Basic understanding of AI; What AI can do and can not do | | | Critical appraisal of clinical AI tools, including the specific use cases and how to use the tools | Understand and interpret the AI-tool results and be able to explain to patients | | Side effects of the tool; Recognizing the ”black-box” nature of AI |
| (Ötleş et al., 2022 [35]); Commentary | Essential AI concepts; Foundational knowledge to AI algorithms | | | Clinical application scenarios of different AI concepts | Statistical tests | | Accountability; Transparency; Fairness/Algorithmic discrimination |

Continued on next page

Table 1: What should medical students know about AI? Existing AI topics proposed to teach (Continued)

| | | | | | | | |
|---|---|---|---|---|---|---|---|
| (Tejani et al., 2022 [47]), Research | Common AI terminologies; Basic understanding of ML principle; | Overview of the breadth of use cases with a discussion of sub-fields within AI, such as computer vision and natural language processing | Health data collection and curation | Fundamental knowledge of translating AI concepts into clinical practice; How AI tools are integrated into current clinical workflow | | | Health data privacy; Safety, and bias in AI models; The legal framework and regulatory compliance of AI-based clinical applications; AI risks to both patients and providers |
| (Kalthoff et al., 2022 [16]), Research | Basic understanding of AI definition and common terminologies; Symbolic and subsymbolic methods of AI; How AI can be used by physicians | | Health data management, curation, protection and security | Different application scenarios of AI in medicine; Potential and importance of AI for different occupational fields in health system; Future prospects and development of AI research in health sector | | | Ethical and social implications; Legal aspects |
| (Waldman et al., 2022 [50]), Review | Basic understanding of AI; Common ML algorithms | Basic understanding of computer science; Basic understanding of deep learning (e.g., neural networks) | Big data in healthcare; How to prepare health data for ML algorithm inputs | Clinical utility of ML, such as precision medicine; | Statistical analysis methods; Receiver operator analysis | Design a ML algorithm | Black-box effect; Bias and explainability; Reproducibility |
| (Alam et al., 2023 [3]), Opinion | Common AI terminologies; Basic understanding of AI | | | Common clinical application scenarios of generative AI; Think independently and appraise critically for responsible use of medical AI tools | | | Accountability in making clinical decisions. |
| (Ng et al., 2023 [32]), Perspective | Basic understanding of AI; Basic ML algorithms | Basic understanding of neural networks and deep learning | Health dataset curation process | | Confusion matrices; Sensitivity, specificity, accuracy, predictive values; Statistical tests | | Accountability; Bias and explainability; Informed Consent |

Continued on next page

Table 1: What should medical students know about AI? Existing AI topics proposed to teach (Continued)

| | | | | | | | |
|---|---|---|---|---|---|---|---|
| (Weidener et al., 2023 [52]), Research | Basic understanding of AI | Basic concepts of statistics | | Practical skills of using AI-based tools; Trust and confidence of using AI-based tools | Interpreting the results of AI applications (explainability); Evaluating the results of AI applications (statistical analysis) | | Ethics; Data protection and privacy; Data bias; Associated risks |
| (Weidener et al., 2024 [53]), Viewpoint | | | | | | | Informed consent; Bias; Safety; Transparency; Privacy; Allocation; Fairness; Responsibility; Empathy; Explainability; Liability; Accountability |

Table 2: Practical teaching efforts providing AI training to medical students

| Author / Year / Country | Year of Implementation | Duration | Type | Number of Participants | Structure and Teaching Materials | AI Topics Covered | Instructor Expertise | Education Tools | Assessment | Key Findings / Outcomes | Main Limitations |
|---|---|---|---|---|---|---|---|---|---|---|---|
| (Lang et al., 2020 [21]), Germany | 2017-19 | 1 semester | Elective Course | 47 | 11 sessions (2 hours/each). Each session with demonstration, concrete development of algorithms, and guided discussions | AI in personal everyday life; AI examples in medicine; Scientific use and potentials of AI; Neural networks and principles of deep learning; Basic knowledge of statistics; Mathematical description in AI procedures; Basic programming and practical implementation. Ethics and responsibility, limitations, and possible. | Experienced AI users | | Pre- and post study survey about AI knowledge and Interest in AI topics | Self-assessed knowledge increased from an initial 2.3 to 3.8 on a 5-point Likert scale; Interest in the AI topic was initially 4.3 and increased to 4.8. The interest of medical students in AI could be significantly increased by a structured integration in the curriculum | |
| (Werner et al., 2020 [54]), Germany | 2019 | 2 weeks | Mandatory Course + Elective Course | 13 | Integrated into medical degree program consisting of a modularly structured core curriculum and elective courses | Telemedicine; Robotics, automation and virtual/augmented reality in medicine; AI; Big data: Omics and biomarkers for personalized medicine; Smart medical devices health apps | Faculties from departments of Medicine, Informatics, Physics and Law | | Post study questionnaire-based evaluation about overall satisfaction after each module | Students' assessment in the final feedback round was consistently very positive, suggesting that the presented concept is in general attractive | |

Continued on next page

Table 2: Practical teaching efforts providing AI training to medical students (Continued)

| | | | | | | | | | | |
|---|---|---|---|---|---|---|---|---|---|---|
| (Sendak et al., 2021 [45]), United States | 2016-2019 | 1 year | Extra-curricul Program | 14 m | Online course, supplemented with on site programs and hands on experience; community discussion | Innovations in medical AI; Value based care; Policy regarding AI-based medical devices; AI algorithms, Bias; Regulatory Compliance | | | Pre- and post-study survey about the AI understanding on a 5 point scale (1 = limited understanding; 5 = deep understanding) | Students' understanding across all domains increased significantly; 7 out of 11 domains demonstrating improvements in understanding AI. | Small cohort, findings may not be generalizable to other institutions; no control group in this study to compare outcomes |
| (Lee et al., 2021 [23]), Italy | 2020 | 7 weeks | Extra-curricul Program | 45 m | Series of lectures and interactive sessions (total 12 sessions, 2 hours/each); 3 days of live workshops with project-based learning with a competitive task to develop an AI tool | Medical AI engineering; A competitive task to develop an medical AI application; AI Ethics and regulatory policy | | | Pre- and post-study survey about AI competencies; Post-study survey about overall course satisfaction | Overall satisfaction with the AI course was high Students' self-assessed competency in AI showed a striking increase in all AI learning out-comes | Time differences hindering simultaneous gathering of all students; the learning needs from different cultures/systems, human interactions could not be fully satisfied |

Continued on next page

| | | | | | | | | | | | |
|---|---|---|---|---|---|---|---|---|---|---|---|
| (Laupichler et al., 2022 [22]), Germany | 2021-2022 | 7 months | Elective Course | 24 | Online course with self-study content as one main component, supplemented with community discussions | How AI is supporting the diagnostic process using medical image data; Opportunities and risks of medical AI application; The future of physicians in the context of AI. | AI researchers | MOOC platform "AI-Campus"; web-based DICOM viewer to view medical imaging data | Pre and post-study assessment of "medical AI readiness scale for medical students (MAIRS-MS)" | Statistically significant increase in medical students' AI readiness | Using performance tests (e.g., knowledge tests) rather than self-assessment questionnaires as MAIRS-MS test focusing on attitudes and affect |

Continued on next page

Table 2: Practical teaching efforts providing AI training to medical students (Continued)

| | | | | | | | | | | | |
|---|---|---|---|---|---|---|---|---|---|---|---|
| (Krive et al., 2023 [19]), United States | 2019 | 4 weeks | Elective Course | 20 | Foundational readings; live online lectures; collaboration projects in small groups; community discussion | Role of clinicians in future AI-powered digitally enabled workplace; Common AI terminologies; Applications of AI in patient care; Appraise the validity and relevance of articles as input in AI models; Team projects integrating and applying AI using real-world evidence and medication safety data; AI for efficient and effective healthcare delivery; Ethics and regulatory, legal issues framing AI; Bias in data and implications for AI models | Faculties from diverse background including AI, medicine, and instructional design | An online learning management system | Pre- and post-study assessment of AI knowledge; Post-study interview | Students demonstrated an improvement in knowledge from an average quiz score of 68% to 97%; Students indicated that they learned how to transition from the theory to the practice of AI Students stated they developed new perspectives regarding the roles of AI. Students developed alternative viewpoints, learned that the importance of the construct of quality measures and projects | Being an elective, the students who enrolled were naturally interested in this topic area, so the results may not be generalizable to all medical students |
| (Tsopra et al., 2023 [49]), France | 2021-2022 | 1 semester | Elective Course | 15 | 5 sessions (3 hours/each) with interactive lectures and practical projects | AI-clinical decision support system (CDSS) architecture; AI-CDSS for diagnosis; AI-CDSS for therapeutics; Usability and Adoption, Advantages/Limits, Certification/ Evaluation and Safety | A medical informatics teacher wit several clinicians | Interactive platforms: Moodle, Wooclap | Post-study survey of students opinions and feedback (5-point Likert scale) regarding lesson content, lesson format, modalities of examination, global experience, and interests in medical informatics | Overall the programs aroused great enthusiasm and strong engagement of students | Only elective, leading to small sample size; Completely in-person, inflexible; instructors with skills in both clinical medicine and medical informatics are rare; more extensive evaluation can be done |

Continued on next page

Table 2: Practical teaching efforts providing AI training to medical students (Continued)

| | | | | | | | | | | | |
|---|---|---|---|---|---|---|---|---|---|---|---|
| (Pizzolla et al., 2023 [40]), Belgium | 2022 | 16 hours | Mandat Course | ry25 | Online recorded lecture and supplementary reading materials | Introduction to AI, ML, and expert systems; Introduction to ML in healthcare; Introduction to computer vision; Introduction to image recognition in healthcare; Introduction to electronic health records (EHRs) and their role in modern healthcare; The ways in which digital medicine is transforming healthcare delivery; The impact of digital medicine on research and innovation within the medical field; Various applications of AI in healthcare | | | Post-study survey assessing students' perception of the AI program | 45% of respondents rated it as good, 40% rated it as fair, and 15% rated it as poor; AI program generally meets the needs of the participants, but there might be some areas for improving, particularly in addressing AI legal and ethical concerns; AI program should place a greater emphasis on developing students' practical skills in applying AI to medical practice | Low sample size and limited outreach of the program (only one class, in one university, in Belgium); not performing a scoping review of active AI programs in healthcare curricula; not performing any statistical inference on the sample, which largely limits the information gathered from the questionnaire |

Continued on next page

Table 2: Practical teaching efforts providing AI training to medical students (Continued)

| | | | | | | | | | | |
|---|---|---|---|---|---|---|---|---|---|---|
| (Abid et al., 2024 [1]), United States | 2021 | 1 month | Elective Course | 19 | Offered in 2 tracks: Technical and Non-technical. Self-guided online lectures; project-based deliverables, supported by the course's faculty advisor | Introduction to AI and ML; Various ML techniques (supervised/unsupervised, classification/regression, etc); The impact of ML in medicine, broadly and in the student's chosen specialty; Basics of neural networks; Limitations and pitfalls of ML (reproducibility, interpretability, and bias); Issues that may arise in the implementation of an ML algorithm in clinical practice; Ethical issues that concern the use of ML in health care. | A physician with substantial leadership and experience in AI and ML research | | Pre- and post-study assessment about students' confidence in AI and ML skills; Post-study interview about students' experiences | Students' self-reported confidence scores for AI and ML increased by 66%. In qualitative surveys, students reported enthusiasm and satisfaction with the course and commented that the self-direction and flexibility and the project-based design of the course were essential | Small number of participants, especially in the Technical track, restricting the generalizability of this study; Assessments relied only on students' self-reported confidence, which has been shown to be a flawed metric |
| (Kroplin et al., 2024 [20]), Germany | 2022-2023 | 1 semester | Elective Course | 20 | Four AI Topics, each with 3 lessons (90 minutes/each) | Applications of surgical robots; Precision medicine; the potential use of generative AI in teaching, research, and patient treatment | AI researchers | | Pre- and post-study assessment about students' knowledge on AI and digital health | The pre-post test comparison revealed a significant increase from 4/10 to 8.3/10 in knowledge (P.001) among the participants. | Single-center design and the small number of participants; Final test only examing excerpts from topics that cannot represent the full scope of the curriculum |

Continued on next page

Table 2: Practical teaching efforts providing AI training to medical students (Continued)

| | | | | | | | | | | |
|---|---|---|---|---|---|---|---|---|---|---|
| (Park et al., 2024 [38]), United States | 2022-2023 | 1 semester | Elective Course | 10 | An introductory lecture to describe the history and current state of AI in medicine; Four learning objectives with each consisting of three modules: a journal club, a coding demonstration, and an integrative lecture-discussion. | Basic ML algorithms, Basics in neural networks; Health data engineering; ML model construction and validation, and model interpretability | AI researchers | | Pre- and post-survey asking students to rate their own confidence in describing learning objectives; post-survey for students' overall satisfaction | Significant increases in students' self-reported confidence in describing the content objective; students maintained a high degree of interest in data science and AI-focused education. | Heterogeneity of student attendance due to the nature of the elective model; lack of hands-on project building experience that requires AI model building and testing |

## Discussion

This scoping review examines existing recommendations and current clinical education providing AI training and education to medical students.

**Discrepancy between proposed frameworks and current AI education programs for medical students** Published theoretical guidelines emphasize teaching applications of AI and its ethical use [54, 49], yet prevailing AI training programs instead focus on mathematical principles, data science, and coding skills. These areas, while important, may not be directly relevant for medical students and may become less relevant as models become more accessible to lay users. To address this disparity, we propose integrating clinical case studies that highlight the application of AI in diagnostics, treatment planning, and patient monitoring. Similarly, we advocate for workshops and simulations that mimic real-life scenarios [40].

**Challenges of integrating AI training into the medical education program** Despite the calls to introduce AI concepts to medical students [27, 1, 30] and a strong willingness for its study [13, 51, 41], integrating AI training into current medical education programs is difficult and may face resistance. The foremost criticism of incorporating AI into medical curricula is that it has yet to prove its worth to medicine [26]. Theoretical use cases are plentiful for clinical decision support tasks, but these rarely provide information an astute clinician does not already know [55]. Furthermore, while randomized controlled trials show promise for these technologies, their numbers are few and often of questionable generalizability [5]. As with any new medical paradigm, more evidence may be needed to demonstrate how AI-guided clinical decision making improves clinician performance or patient outcomes prior to incorporation into curricula. Secondly, there is the challenge of identifying qualified educators. Few clinicians possess training in the principles of mathematics, computer engineering, or data science, fewer still have experience in developing or implementing clinical models [10]. Education in this space must therefore rely on a collaborative teaching environment and crosstalk between university departments when available. Thirdly, determining the content to teach poses a significant challenge. The field is rapidly evolving, and standardized coursework may quickly become outdated [57]. Medical curricula are undergoing transitions away from traditional lecture and basic science frameworks to case-based learning, clinical reasoning assessments, expansion of social science topics, and earlier exposure to clinical care. While these new curricula are perhaps more adaptable to new topics (i.e. COVID-19 [11]), their true impact on resident and physician performance cannot be measured until years after their implementation. This will also be true of any curriculum changes to incorporate AI.

**The AI literacy framework for medical students** Existing proposals require optimization in terms of (1) Comprehensiveness, or needing to focus on the multiple aspects of AI [53, 31]; (2) Clinical Relevance, or grounding AI in the clinical workflow [29, 45]; and (3) Personalization, or tailoring for each stage of medical education and future career trajectories [27, 23]. Inspired by the cognitive domains in Bloom's taxonomy [5] and how previous researchers categorized AI literacy [31], we propose the framework in Figure 3.

Our framework categorizes medical students' AI literacy into four dimensions: (1) **Foundational AI**: general understanding of AI capabilities and limitations in medicine. (2) **Practical AI**: the ability to select and use the appropriate AI tools for different settings. (3) **Experimental AI**: deeper understanding of evaluation metrics and technical differences between models. (4) **Ethical AI**: how AI can ensure, or detract from patient privacy, equity, autonomy, and well-being.

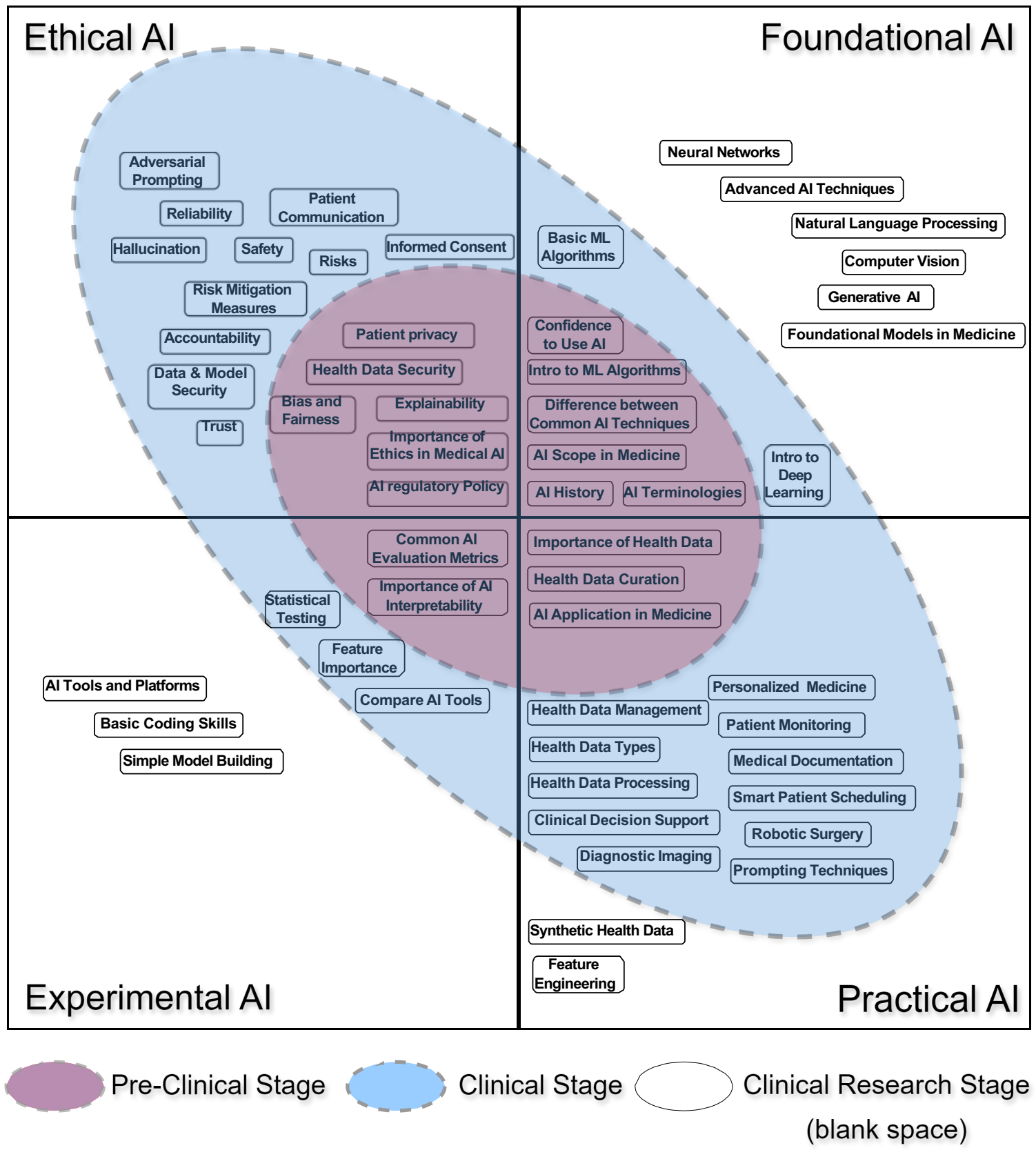

Figure 3. Medical students AI Literacy framework

Our framework further outlines the learning objectives for various stages (i.e., pre-clinical stage, clinical stage, and clinical research stage) of medical students within each of the four dimensions. During the pre-clinical stage, medical students focus on learning basic sciences and doctoring skills. To complement this, our framework addresses misconceptions and builds confidence in the use of AI to enhance clinical learning and practice. Main topics are shown in the inner circle of Figure Figure 3.

In the clinical stage, as medical students transition to patient care, our framework emphasizes competencies tailored to various clinical settings, including workflow of clinical decision support systems, AI-supported diagnostic imaging, multi-omic personalized medicine, patient monitoring, robotic surgery, and streamlining medical documentation. In addition, students need to have solid knowledge in AI safety, assessing risks, managing liability, ensuring legal compliance, maintaining accountability, securing health data and models, and understanding hallucination.

In the clinical research stages, medical trainees or physicians may help implement, develop, or customized existing AI tools. This requires continual education to keep abreast of developments in the AI field.

**Lack of educational tools to support AI learning** Only three out of 11 studies indicated the use of online educational tools, tools which primarily cater to the general public (e.g., MOOCs), thus placing a strong emphasis on foundational AI concepts and coding skills without clinical relevance. Yet, online tools have demonstrated promising outcomes [39, 56] provided flexibility in self-guided learning, and ensured equitable access to those with internet [7]. Having an online resource for medical AI education may offer a highly scalable, upgradeable, equitable, and engaging resource for medical students.

**Limitations of this study** There are several limitations to this study that should be acknowledged. First, the AI literacy framework proposed in this paper partly originates from the less well-founded viewpoints and commentaries, thus introducing subjectivity and bias. Second, the review was conducted up until May 2024, and, given the rapidly evolving nature of AI technologies, it is crucial that the proposed framework be continuously updated to incorporate the latest advancements in the field. Third, this study primarily focuses on medical students. This narrow focus may limit the applicability of the findings to other stages of clinical training as needs evolve. Finally, this systematic review only included papers that were accessible in English and explicitly stated their focus on AI education for medical students. This language and scope restriction may have led to the exclusion of relevant studies published in other languages or with a broader focus.

## Conclusion

Medical students will play a crucial role in the future utilization of medical AI technology, and they need to be adequately equipped to use it safely, ethically, and effectively. Despite continuous calls for introducing AI concepts to medical students, a radical medical educational reform to accommodate AI training is not realistic due to numerous challenges identified in this study. Instead, our proposed AI literacy framework provides a road map for developing a more practical and relevant education strategy tailored to medical trainees. By addressing discrepancy between current theories and practices, education researchers and practitioners can enhance AI education for medical students, ultimately preparing them to effectively use this new tool in their future medical practice.

## Competing interests

The authors declare no competing interests.

## Author contributions

Y.M., A.B. and B.S. conceived the study idea and developed the research questions. Y.M. and J.B. designed the search strategy and developed the inclusion and exclusion criteria. Y.M., Y.S., and S.K. conducted the literature search and screened the identified studies. Data extraction was performed by Y.M., Y.S., D.V., and S.K.. Quality assessment of the included studies was conducted by Y.M. and Y.S.. Y.M. and Y.S. analyzed the data and synthesized the results. Y.M., Y.S., and J.B. developed the AI literacy framework. The manuscript was drafted by Y.M., Y.S., J.B., and D.V., and revised by all other authors. Y.M., Y.S., J.B., Y.R.,

M.B., P.R., T. L., A.B., and B.S. contributed to the interpretation of findings. All authors provided critical feedback and approved the final version for submission.

## Supplementary information

Data supporting this scoping review are available in the original publications, reports, and preprints that were cited in the reference section. In addition, the analyzed data that were used during the review are available from the author on reasonable request.